\documentclass[aps,twocolumn,prb]{revtex4}
\usepackage{graphicx}

\begin{document}


\title{Teaching the hidden symmetry of the Kepler problem in
relativistic quantum mechanics - from Pauli to Dirac electron}



\author{Tamari~T.~Khachidze and Anzor~A.~Khelashvili}
\email[]{khelash@ictsu.tsu.edu.ge}
\affiliation{Department of Theoretical Physics,  Ivane Javakhishvili Tbilisi State University,
                         I.Chavchavadze ave. 3, 0128, Tbilisi, Georgia}


\date{\today}

\begin{abstract}

Hidden symmetry in Coulomb interaction is one of the mysterious problems of modern physics.
Additional conserved quantities associated with extra symmetry govern wide variety of physics problems,
from planetary motion till fine and hyperfine structures of atomic spectra.  In this paper we present a
simple derivation of hidden symmetry operator in relativistic quantum mechanics for the Dirac equation in
the Coulomb field. We established that this operator may be reduced to the one introduced by Johnson
and Lippmann. It is worthwhile to notice that this operator was discussed in literature very rarely and so
is not known well among physicists and was omitted even in the recent textbooks on relativistic quantum
mechanics and/or quantum electrodynamics.

\end{abstract}


\maketitle

\section{Introduction}

It is well known that the Coulomb problem has additional dynamical symmetry in classical and non-relativistic
quantum mechanics. Classical orbits in the Coulomb field are closed due to this symmetry. In quantum mechanics
this symmetry shows up in the form of "accidental" degeneracy of hydrogen atom spectrum. The nature of these
phenomena was explained by V.Fock~\cite{Fock1935}, V. Bargman~\cite{Bargman1936}, W. Pauli~\cite{Pauli26} and
others many years ago. The "hidden" symmetry related applications have been discussed in several
textbooks~\cite{Goldstein2002,Jose1998,Bohm1986} as well. The conserved quantity associated with this symmetry
often called the Laplace-Runge-Lentz (LRL) vector  were discussed by many authors (see~\cite{Ross2003,Valent2003}, for instance).

Less information is published about this symmetry in the relativistic quantum mechanics, in particular, for the Dirac
equation. It is known, that well known $l$-degeneracy of hydrogen atom spectrum in non-relativistic quantum mechanics,
is removed in case of the Dirac equation (since LRL vector is no longer conserved in this case). However the degeneracy
is not lifted completely even in this case. The two-fold degeneracy in total angular momentum still remains and it
contradicts with existence of well known experimental result - the Lamb shift. It is entirely possible that the Lamb
shift is also a consequence of the same symmetry and it was discussed in this content in relatively recent
papers~\cite{Tangerman93,Sukumar1985}.

The generalization of LRL vector that involves spin degrees of freedom (Dirac's particle) is called Johnson-Lippman (JL) operator.
The aim of our paper is to derive the JL operator for relativistic hydrogen atom in the Dirac equation as easily as possible.

\section{The Johnson-Lippman operator (A short history)}

The analogue of LRL vector in the Dirac equation is the Johnson-Lippmann (JL) operator which in $1950$ was published by these authors
in the form of brief abstract~\cite{Johnson1950}. Because of importance and rarity of this problem, below we display the full text of
the above mentioned abstract as it was presented then.

"\textbf{J6. relativistic Kepler Problem,} M.H. Johnson and B.A. Lippman, \textit{Naval Research Laboratory} - Besides the usual
integrals of motion, $\vec M$ and $\vec j$ (in Dirac's notation) the relativistic equations for a charge in a Coulomb field admit
\begin{equation}\label{E:JohnsonLippman}
    A=\vec \sigma \vec rr^{-1}-i\left({\hbar c\over e^{2}}\right)(mc^{2})^{-1}j\rho_1 (H-mc^{2}\rho _{3})
\end{equation}
as another integral of motion. Since $A$ and $j$ anticommute, the pairs with the same $|j|$ are degenerate. Thus the existence of
$A$ establishes the "accidental" degeneracy in relativistic Kepler problem just as the existence of the axial vector establishes
the degeneracy with respect to $l$ in the corresponding non-relativistic problem".

It is all what is known about this operator in scientific press and textbooks. More detailed consideration, in our best knowledge,
had not been published neither then nor after. As far as textbooks concerned, we have found only a brief remark in the form of
a short footnote in B.V. Berestetskii's book~\cite{Berestetskii1982}.

Therefore we think that the derivation of JL operator and consideration its principal properties is worthy of certain interest for
students as well as for teachers of universities.

The JL operator were generalized to arbitrary dimensions in the recent paper~\cite{Katsura2004} and it was shown that this operator
can be used to construct relativistic supercharge in for so-called Witten's superalgebra. As far as commutativity of JL operator
with the Dirac Hamiltonian is concerned, it was only mentioned that it can be proved by "rather tedious manipulations".

We think that from pedagogic point of view it is important to show explicitly the simplest way to prove the above statement. Below
we construct this operator in rather simple and very transparent and understandable way.

\section{Dirac's $K$-operator and $K$-odd operators}

It is well known that the Dirac's operator
\begin{equation}\label{E:DiracOperator}
    K=\beta (\vec \Sigma \cdot \vec l+1)
\end{equation}
commutes with the Dirac Hamiltonian
\begin{equation}\label{E:DiracHamiltonian}
    H=\vec \alpha \cdot \vec p+\beta m-{\alpha \over r},\  \  \  \  a \equiv Ze^{2}=Z\alpha
\end{equation}
where $\vec l$ is the angular momentum vector. Here $\vec \alpha $ and $\vec \beta $ are usual Dirac matrices and $e^{2}=\alpha $ is
is the fine structure constant, while $\vec \Sigma $ is the electron spin matrix
\begin{equation}\label{E:SpinMatrix}
    \vec \Sigma =\rho _{1}\vec \alpha =\gamma ^{5}\vec \alpha =\pmatrix{ \vec\sigma & 0\cr 0 & \vec \sigma }
\end{equation}
According Sommerfeld formula the hydrogen atom spectrum is given as follows~\cite{Sommerfeld209}:
\begin{equation}\label{E:SommHSpectra}
    {E\over m}=\left\{ 1+{(Z\alpha )^{2}\over \left(n-|\kappa |+\sqrt {\kappa ^{2}-(Z\alpha )^{2}}\right)^{2}}\right\} ^{-1/2}
\end{equation}
where $\kappa $ is the eigenvalue of a $K$, $|\kappa |=\sqrt {j(j+1)+1/4}=j+1/2.$ Since $\kappa $ can have two signs the energy spectrum
is two-fold degenerate. In equation (\ref{E:JohnsonLippman}) for JL operator the Dirac operator is denoted by $j$. In our approach we
follow the simple logic:

-   If there is some symmetry that relates two signs of $\kappa$,
the corresponding symmetry operator must anticommute  with $K$.

-   At the same time, naturally, this operator has to commute with the Dirac Hamiltonian.

Thus the first step will be constructing an operator(s), that
anticommute with $K$. In order to find such an operator, let
generalize the theorem, which was known earlier for the Pauli
equation. We reformulate it in the following form. \newline{\bf
Theorem:}
    Let $\vec V$ be a vector with respect to the angular momentum  $\vec{l}$, i.e.,
    $$\left[l_{i},V_{j}\right]=i\varepsilon _{ijk}V_{k}$$
    In the vector product form it can be written as
    $$\vec l\times \vec V+\vec V\times \vec l=2i\vec V$$
    Suppose also that this vector is perpendicular to  $\vec{l}$
    $$\left(\vec l\cdot \vec V\right)=\left(\vec V\cdot \vec l\right)=0$$
    Then $K$ anticommutes with operator $\left(\vec \Sigma \cdot \vec V\right)$, which is scalar with respect to the total $\vec{J}$ momentum.

{\bf Proof:} Let us consider a product $\left(\vec \Sigma \cdot
\vec l\right)\left(\vec \Sigma \cdot \vec V\right)$. Exploiting
the known properties of Dirac matrices and conditions of the
theorem, one can establish that $$\left(\vec \Sigma \cdot \vec
l\right)\left(\vec \Sigma \cdot \vec V\right)=\left(\vec l\cdot
\vec V\right)+i\left(\vec \Sigma ,\vec l\times \vec V\right)=$$ $$
=i\left(\vec \Sigma ,2i\vec V-\vec V\times \vec
l\right)=-2\left(\vec \Sigma\cdot \vec V\right)-i\left(\vec \Sigma
,\vec l\times \vec V\right).$$ Therefore
\begin{equation}\label{E:Theorem1}
    \left(\vec \Sigma \cdot \vec l+1\right)\left(\vec \Sigma \cdot \vec V\right)=-\left\{ \vec \Sigma \cdot \vec V+i\left(\vec \Sigma ,\vec l\times \vec V\right)\right\}.
\end{equation}
Now consider the same product in reversed order
$$\left(\vec \Sigma \cdot \vec V\right)\left(\vec \Sigma \cdot \vec l\right)=\left(\vec V\cdot \vec l\right)+i\left(\vec \Sigma ,\vec V\times \vec
l\right)=i\left(\vec \Sigma ,\vec V\times \vec l\right).$$
Hence
$$\left(\vec \Sigma \cdot \vec V\right)\left(\vec \Sigma \cdot \vec l+1\right)=\left(\vec \Sigma \cdot \vec V\right)+i\left(\vec \Sigma ,\vec V\times \vec l\right)=$$
$$ =-\left(\vec \Sigma \cdot \vec l+1\right)\left(\vec \Sigma \cdot \vec V\right).$$
In the last step we made use of the equation (\ref{E:Theorem1}). Therefore we have obtained
\begin{equation}\label{E:Theorem2}
    \left(\vec \Sigma \cdot \vec l+1,\vec \Sigma \cdot \vec V\right)=0
\end{equation}
Now, according to the definition (\ref{E:DiracOperator}), it
follows that
\begin{equation}\label{E:Theorem3}
    K\left(\vec \Sigma \cdot \vec V\right)=-\left(\vec \Sigma \cdot \vec V\right)K
\end{equation}
{\bf Thus the theorem is proved.}

It is evident that the class of anticommuting with $K$ (so called, $K$ -odd) operators is not restricted by these operators only - any
operator of kind $\hat O\left(\vec \Sigma \cdot \vec V\right)$ , where  $\hat O$ is commuting with $K$, but otherwise arbitrary, also is $K$ -odd.

It is noteworthy, that the following useful relation holds in the framework of constraints of above theorem:
\begin{equation}\label{E:UsefulRel}
    K\left(\vec \Sigma \cdot \vec V\right)=-i\beta \left(\vec \Sigma ,{1\over 2}\left[\vec V\times \vec l-\vec l\times \vec V\right]\right),
\end{equation}
which follows from vector's transformation rule mentioned in theorem in the form of vector product.

One can see that the antisymmetrized vector product characteristic of LRL vector appears on the right-hand side of this relation. Important
special cases, resulting from the above theorem include $\vec V=\hat {\vec r}$ (unit radial vector), $\vec V=\vec p$ (linear momentum) and
$\vec V=\vec A$ (LRL vector). This last one has the following form\cite{Pauli26}
\begin{equation}\label{E:AVector}
    \vec A=\hat{\vec{r}}-{i\over 2ma}\left[\vec p\times \vec l-\vec l\times \vec p\right]
\end{equation}

According to (\ref{E:UsefulRel}) there is a relation between above three odd operators
\begin{equation}\label{E:ASigma}
    \vec \Sigma \cdot \vec A=\vec \Sigma \cdot \hat {\vec r}+{i\over ma}\beta K\left(\vec \Sigma \cdot \vec p\right)
\end{equation}
At this point we are tooled up to derive the hidden symmetry operator, which is done in the following section.

\section{The hidden symmetry operator}

The second step of our two stage derivation strategy is to find  $K$-odd operators that commutes with the Dirac Hamiltonian.
There remains still considerable freedom of choice here because of above mentioned remark about operators like $\hat O\left(\vec \Sigma \cdot \vec V\right)$.
One can take $\hat O$ into consideration or ignore it.

Let us choose
\begin{equation}\label{E:12Twelve}
    \vec \Sigma \cdot \hat {\vec r}\  \  \  and\  \  K\left(\vec \Sigma \cdot \vec p\right)
\end{equation}
Because of specific place of LRL vector, it is reasonable to
assume that $\vec \Sigma \cdot \vec A$ - like term will present
here. So this choice is dictated by (\ref{E:ASigma}). Both
operators in (\ref{E:12Twelve}) are diagonal matrices. After
commuting them with non-diagonal $H$  we will eventually end up
with non-diagonal terms. For example, $$\left[\vec \Sigma \cdot
\hat {\vec r},H\right]={2i\over r}\beta K\gamma ^{5}$$ (See,
exercise $2$ below). The resulting matrix on the right-hand side
is anti-diagonal. Probing the following combination of diagonal
and anti-diagonal matrices yields:
\begin{equation}\label{E:APrimeMat}
    A'=x_{1}\left(\vec \Sigma \cdot \hat {\vec r}\right)+ix_{2}K\left(\vec \Sigma\cdot \vec p\right)+ix_{3}K\gamma ^{5}f(r)
\end{equation}

Here the coefficients are chosen in such a way, that $A'$ be Hermitian, with $x_{1,2,3}$ are arbitrary real numbers and $f(r)$ is an arbitrary scalar function
(to be determined later).

The commutator of $A'$ with $H$ is given by
\begin{eqnarray}\label{E:APrimeHComm}
    [A',H]=x_{1}{2i\over r}\beta K\gamma ^{5}-x_{2}{a\over r^{2}}K\left(\vec \Sigma \cdot \hat {\vec r}\right)-\nonumber\\
    -x_{3}f'(r)K\left(\vec \Sigma \cdot \hat {\vec r}\right)-ix_{3}2m\beta K\gamma^{5}f(r)
\end{eqnarray}
Grouping diagonal and antidiagonal matrices separately and equating this expression to zero, we obtain equation
\begin{eqnarray}\label{E:AntiDiagonal}
    K\left(\vec \Sigma \cdot \hat {\vec r}\right)\left({a\over r^{2}}x_{2}+x_{3}f'(r)\right)+\nonumber\\
    +2i\beta K\gamma ^{5}\left({1\over r}x_{1}-mf(r)x_{3}\right)=0
\end{eqnarray}
This equation is satisfied, if diagonal and antidiagonal terms become zero separately, i.e.,
\begin{equation}\label{E:Condition}
    {1\over r}x_{1}=mf(r)x_{3},\  \  \  \  \  \  {a\over r^{2}}x_{2}=-f'(r)x_{3}
\end{equation}
Integration of the second equation over the interval $(r,\infty )$ yelds
\begin{equation}\label{E:Integration}
    x_{3}f(r)=-{a\over r}x_{2}
\end{equation}
and using this result in the first equation, we obtain that
\begin{equation}\label{E:Result1}
    x_{2}=-{1\over ma}x_{1}
\end{equation}
\begin{equation}\label{E:Result2}
    x_{3}f(r)={1\over mr}x_{1}
\end{equation}
Substitution of (\ref{E:Integration}), (\ref{E:Result1}) and (\ref{E:Result2}) into (\ref{E:APrimeMat}) yields the following operator that commutes
with the Dirac Hamiltonian, (\ref{E:DiracHamiltonian}):
\begin{equation}\label{E:DirHam}
    A'=x_{1}\left\{ \left(\vec \Sigma \cdot \hat {\vec r}\right)-{i\over ma}K\left(\vec \Sigma \cdot \vec p\right)+{i\over mr}K\gamma ^{5}\right\}
\end{equation}
This operator is a $K$ -odd and satisfies all conditions of the above theorem.

Thus we constructed the operator, which is associated with a hidden symmetry of the Dirac equation in Coulomb field. But this operator is not new.
One can make sure of that (\ref{E:DirHam}) is a different form of JL operator. Indeed, if we turn to usual $\vec \alpha $  matrices making use of
definition (\ref{E:SpinMatrix}) and taking into account the expression (\ref{E:DiracHamiltonian}) for the Dirac Hamiltonian, $A'$ can be reduced to
the form  ($x_{1}$, as unessential common factor, may be dropped)
$$
    A'=\gamma ^{5}\left\{ \vec \alpha \cdot \hat {\vec r}-{i\over ma}K\gamma ^{5}\left(H-\beta m\right)\right\}
$$
This expression is nothing but the Johnson-Lippmann operator, (\ref{E:JohnsonLippman}).

\section{Physical Meaning and some applications of JL operator}

In order to understand a physical meaning of $A'$ operator note that, using (\ref{E:UsefulRel}), equation (\ref{E:DirHam}) may be rewritten in the following form:
\begin{equation}\label{E:APrime21}
    A'=\vec \Sigma \cdot \left(\hat {\vec r}-{i\over 2ma}\beta \left[\vec p\times \vec l-\vec l\times \vec p\right]\right)+{i\over mr}K\gamma ^{5}
\end{equation}
In the non-relativistic limit, when $\beta \rightarrow 1$, and $\gamma ^{5}\rightarrow 0$, this operator reduces to the projection of LRL vector on the
electron spin direction, $A'\rightarrow \vec \Sigma \cdot \vec A$, or because of $\left(\vec l\cdot \vec A\right)=0$, it is a projection on the
total momentum, $\vec J$.

As for further application, let us calculate the square of JL
operator. The result is as follows~\cite{Katsura2004}:
\begin{equation}\label{E:JLSquared}
    A^{2}=1+\left({K\over a}\right)^{2}\left({H^{2}\over m^{2}}-1\right)
\end{equation}
Because all operators in (\ref{E:JLSquared}) commute with each other, one can replace them by their eigenvalues. Therefore one obtains energy spectrum pure
algebraically after specifying spectrum of $A^{2}$. Since $A^{2}$ is positively defined the minimal eigenvalue of $A^{2}$ is zero. For this eigenvalue
solution of (\ref{E:JLSquared}) gives precisely the ground state energy of hydrogen atom,
$$
    E_{0}=m\left(1-{\left(Z\alpha \right)^{2}\over \kappa ^{2}}\right)^{1/2}
$$
Full spectrum can be easily derived by well-known ladder procedure~\cite{Katsura2004} . It leads to the usual Sommerfeld formula (\ref{E:SommHSpectra}).

It is worthwhile to note a full analogy with classical mechanics, where closed orbits were derived by calculating the square of LRL vector without solving
the differential equation of motion~\cite{Goldstein2002}. Therefore we are convinced that the degeneracy of hydrogen atom spectrum with respect to interchange
$\kappa \rightarrow -\kappa $ is related to the existence of JL operator, which in its turn takes its physical origin from the Laplace-Runge-Lenz vector.

It is also remarkable that the same symmetry is responsible for absence of the Lamb shift in this problem. Inclusion of Lamb shift terms,~\cite{Kaku1993}
\begin{eqnarray}\label{E:LambShift}
    \triangle V_{Lamb}\approx {4\alpha ^{2}\over 3m^{2}}\left(\ln {m\over \mu }-{1\over 5}\right)\delta ^{3}(\vec r)+ \nonumber\\
    +{\alpha ^{2}\over 2\pi m^{2}r^{3}}\left(\vec \Sigma \cdot \vec l\right)
\end{eqnarray}
found by calculating radiative corrections to the photon propagator and photon-electron vertex function, into the Dirac Hamiltonian breaks commutativity
of $A$ with $H$. However it is evident that without radiative corrections, terms like (\ref{E:LambShift}) do not appear in the Dirac Hamiltonian, as in
the one-electron theory, and as long as only Coulomb potential is considered, the appearance of the Lamb shift should be always forbidden.

In conclusion we want to underline once again that the hidden symmetry, associated to the Coulomb potential, governs a wide range of phenomena  from planetary
motion till fine and hyperfine structure of atomic spectra. Therefore teaching of special aspects of this extremely unusual symmetry can extend the capabilities
of advanced students.

Finally we present some useful exercises that can help students to master calculation of commutators, considered in the main text.

\section{Exercises}

Making use of these exercises students will enhance their knowledge in technical problems of manipulations with Dirac matrices and calculations of
some commutators.

\textbf{Exercise 1.}

Show, that the Dirac's operator  commutes with Hamiltonian, i.e. $\left[K,H\right]=0$, for arbitrary central potential, $V(r)$.

\textbf{Solution:}
\begin{eqnarray}
    \left[K,H\right]=\left[\beta \left(\vec \Sigma \cdot \vec l\right)+1,\  \vec\alpha \cdot \vec p+\beta m+V(r)\right]= \nonumber\\
    \left[\beta \left(\vec \Sigma \cdot \vec l\right),\  \vec \alpha \cdot \vec p\right]+\left[\beta ,\  \vec \alpha \cdot \vec p\right]=\beta \left(\vec \Sigma
    \cdot \vec l\right)\left(\vec \alpha \cdot \vec p\right)- \nonumber\\
    -\left(\vec \alpha \cdot \vec p\right)\beta \left(\vec \Sigma \cdot \vec l\right)+2\beta \left(\vec \alpha \cdot \vec p\right)=\beta \left(\vec \Sigma
    \cdot \vec l\right)\left(\vec \alpha \cdot \vec p\right)+ \nonumber\\
    +\beta \left(\vec \alpha \cdot \vec p\right)\left(\vec \Sigma \cdot \vec l\right)+2\beta \left(\vec \alpha \cdot \vec p\right) \nonumber
\end{eqnarray}
perform transitions from alpha to sigma matrices
\begin{eqnarray}
    \left[K,H\right]=\beta \gamma ^{5}\Bigl\{ \left(\vec \Sigma \cdot \vec l\right)\left(\vec \Sigma \cdot \vec p\right)+\left(\vec \Sigma \cdot \vec
    p\right)\left(\vec \Sigma \cdot \vec l\right)+ \nonumber\\
    +2\left(\vec \Sigma \cdot \vec p\right)\Bigr\} =\beta \gamma ^{5}\Bigl\{-2\left(\vec \Sigma \cdot \vec p\right)-i\left(\vec \Sigma ,\vec p\times \vec
    l\right)+ \nonumber\\
    +i\left(\vec \Sigma ,\vec p\times \vec l\right)+2\left(\vec \Sigma \cdot \vec p\right)\Bigr\} =0 \nonumber
\end{eqnarray}
At the last step we made use of theorem given in the text.

In remaining exercises below calculation of commutators are require, that are needed for calculation of commutator of the symmetry operator  with the
Dirac Hamiltonian.

\textbf{Exercise 2}

Calculate the commutator $\left[\vec \Sigma \cdot \hat {\vec r,}H\right]$.

\textbf{Solution:}
\begin{eqnarray}
    \left[\vec \Sigma \cdot \hat {\vec r},\vec \alpha \cdot \vec p+\beta m+V(r)\right]=\left[\vec \Sigma \cdot \hat {\vec r},\vec \alpha \cdot \vec p\right]= \nonumber\\
    \Sigma _{i}\left[{r_{i}\over r},\alpha _{j}p_{j}\right]+ +\left[\Sigma _{i},\alpha _{j}p_{j}\right]{r_{i}\over r} \nonumber
\end{eqnarray}
Let calculate individual terms:
\begin{eqnarray}
    \Sigma _{i}\left[{r_{i}\over r},\alpha _{j}p_{j}\right]=\Sigma _{i}{1\over r}\left[r_{i,}\alpha _{j}p_{j}\right]+\Sigma _{i}\left[{1\over r}
    ,\alpha _{j}p_{j}\right]r_{i}= \nonumber\\
    {\Sigma _{i}\alpha _{j}\over r}\left[r_{i},p_{j}\right]+\Sigma _{i}\alpha _{j}\left[{1\over r},p_{j}\right]r_{i}={i\over r}\vec \Sigma \cdot \vec \alpha
    -i\vec \Sigma \cdot \vec r{\vec \alpha \cdot \vec r\over r^{3}}= \nonumber\\
    ={3i\over r}\gamma ^{5}-{i\over r}\gamma ^{5}={2i\over r}\gamma ^{5} \nonumber
\end{eqnarray}
\begin{eqnarray}
    \left[\Sigma _{i},\alpha _{j}p_{j}\right]{r_{i}\over r}=2\varepsilon _{ijk}\alpha _{k}p_{j}r_{i}\left({1\over r}\right)= \nonumber\\
    =2i\varepsilon _{ijk}\alpha _{k}\left(r_{i}p_{j}-i\delta_{ij}\right)\left({1\over r}\right)= \nonumber\\
    =2i\alpha _{k}l_{k}\left({1\over r}\right)={2i\over r}\gamma ^{5}\left(\vec \Sigma \cdot \vec l\right) \nonumber
\end{eqnarray}
Grouping these terms together, we derive
\begin{eqnarray}
    \left[\vec \Sigma \cdot \hat {\vec r},H\right]={2i\over r}\gamma ^{5}+{2i\over r}\gamma ^{5}\left(\vec \Sigma \cdot \vec l\right)= \nonumber\\
    ={2i\over r}\gamma ^{5}\left(\vec \Sigma \cdot \vec l+1\right)={2i\over r}\gamma ^{5}\beta K \nonumber
\end{eqnarray}

\textbf{Exercise 3}

Calculate  the commutator $\left[K\left(\vec \Sigma \cdot \vec
p\right),H\right]$

\textbf{Solution:}
\begin{eqnarray}
    \left[K\big(\vec \Sigma \cdot \vec p\big),H\right]=K\left[\vec \Sigma \cdot\vec p,H\right]+\left[K,H\right]\big(\vec \Sigma \cdot \vec p\big)= \nonumber\\
    =K\left[\vec \Sigma \cdot \vec p,\vec \alpha \cdot \vec p+\beta m+V(r)\right]=K\left[\vec \Sigma \cdot \vec p,\vec \alpha \cdot \vec p\right]+ \nonumber\\
    +mK\left[\vec \Sigma \cdot \vec p,\beta \right]+K\left[\vec \Sigma \cdot \vec p,V(r)\right] \nonumber
\end{eqnarray}
Let calculate individual commutators
\begin{eqnarray}
    \left[\vec \Sigma \cdot \vec p,\vec \alpha \cdot \vec p\right]=\left[\Sigma_{i},\alpha _{i}\right]p_{i}p_{j}=2i\varepsilon _{ijk}\alpha _{k}p_{i}p_{j}=0 \nonumber
\end{eqnarray}
\begin{eqnarray}
    \left[\vec \Sigma \cdot \vec p,\beta \right]=0 \nonumber
\end{eqnarray}
\begin{eqnarray}
    \left[\vec \Sigma \cdot \vec p,V(r)\right]=\Sigma_{i}\left[p_{i},V(r)\right]=-i\Sigma _{i}{r_{i}\over r}V'(r)= \nonumber\\
    =-i\left(\vec \Sigma \cdot \hat {\vec r}\right)V'(r) \nonumber
\end{eqnarray}

For the Coulomb potential $V(r)=-a / r \rightarrow V'(r)=a/ r^{3}$
and hence $$ \left[K\left(\vec \Sigma \cdot \vec
p\right),H\right]=-{ia\over r^{3}}\left(\vec \Sigma \cdot \hat
{\vec r}\right)$$

\textbf{Exercise 4}
Calculate  commutator $\left[K\gamma ^{5}f(r),H\right]$

\textbf{Solution:}

\begin{eqnarray}
    \left[K\gamma ^{5}f(r),H\right]=K\left[\gamma^{5}f(r),H\right]+\left[K,H\right]\gamma ^{5}f(r)= \nonumber\\
    =K\gamma ^{5}\left[f(r),H\right]+K\left[\gamma ^{5},H\right]f(r)=K\gamma^{5}\left[f(r),\vec \alpha \cdot \vec p\right]+ \nonumber\\
    +K\cdot 2m\gamma ^{5}\beta f(r)=K\gamma ^{5}i\vec \alpha \cdot \hat {\vec r}f'(r)+2mK\gamma ^{5}\beta f(r)= \nonumber\\
    =iK\left(\vec \Sigma \cdot \hat {\vec r}\right)f'(r)+2mK\gamma ^{5}\beta f(r) \nonumber
\end{eqnarray}

These relations are applied above in calculating of equation (\ref{E:APrimeHComm}).

\section{Acknowledgments}

This work was supported by the NATO Reintegration Grant No. FEL.
REG. $980767$.

\end{document}